\documentclass[10pt]{article}
\usepackage{epsfig}

\newcommand{\sect}[1]{\setcounter{equation}{0}\section{#1}}
\renewcommand{\theequation}{\arabic{section}.\arabic{equation}}

\def\be{\begin{equation}}
\def\ee{\end{equation}}
\def\bea{\begin{eqnarray}}
\def\eea{\end{eqnarray}}

\def\part{\partial}

\def\STr{\mbox{STr}}

\def\incl{{\mbox{\footnotesize i}}}
\def\ii{{\incl_X \incl_X}}

\def\Ortin{Ort{\'\i}n}

\def\mn{{\mu\nu}}

\def\makeatletter{\catcode`\@=11}
\makeatletter
\def\mathbox#1{\hbox{$\m@th#1$}}%
\def\math@ccstyles#1#2#3#4#5#6#7{{\leavevmode
       \setbox0\mathbox{#6#7}%
       \setbox2\mathbox{#4#5}%
       \dimen@ #3%
       \baselineskip\z@\lineskiplimit#1\lineskip\z@
       \vbox{\ialign{##\crcr
              \hfil \kern #2\box2 \hfil\crcr
              \noalign{\kern\dimen@}%
              \hfil\box0\hfil\crcr}}}}
\def\mathaccstyles{\math@ccstyles\maxdimen}
\def\maththroughstyles{\math@ccstyles{-\maxdimen}}
\def\unity%
  {\maththroughstyles{.45\ht0}\z@\displaystyle
{\mathchar"006C}\displaystyle 1}
\begin{document}

\rightline{KUL-TF-06/17}
\rightline{hep-th/0605227}
\rightline{May 2006}
\vspace{1truecm}

\centerline{\LARGE \bf Multiple fundamental strings and waves  }
\vspace{.5truecm}
\centerline{\LARGE \bf to non-linear order in the background fields}
\vspace{1.3truecm}

\centerline{
    {\large \bf Joke Adam}\footnote{E-mail address: 
                                  {\tt joke.adam@fys.kuleuven.be}},
                                                            }
\vspace{.4cm}
\centerline{{\it Instituut voor Theoretische Fysica, K.U. Leuven,}}
\centerline{{\it Celestijnenlaan 200D,  B-3001 Leuven, Belgium}}

\vspace{2truecm}


\centerline{\bf ABSTRACT}
\vspace{.5truecm}
\noindent
The Chern-Simons actions of the multiple fundamental string and the multiple gravitational wave are established to full order in the background fields. Gauge invariance is checked. Special attention is drawn to the non-Abelian gauge transformations of the world-volume fields.

\newpage
\sect{Introduction}\label{sectintro}
A multiple brane consists of $N$ branes lying on top of each other \cite{Witten}. The separation between the branes is of order of the string length. Strings stretching between them give rise to new massless modes, in addition to the modes of the strings going from a brane to itself. These extra modes fill out the $U(1)^N$ symmetry of the separate branes into a $U(N)$ symmetry group, giving the brane-stack a non-Abelian structure. At the worldvolume level the non-Abelian structure is represented by the $N$ Born-Infeld vectors combined into one $U(N)$ vector $V$, and the transverse scalars enhanced to matrix coordinates $X^{\mu}$ transforming under the adjoint of $U(N)$. Background fields are function of the matrix coordinates via a non-Abelian Taylor expansion\cite{Douglas, GM}:
\begin{eqnarray}
C_\mn (x^a, X^i) = \sum_n \frac{1}{n!} \part_{k_1} ... \part_{k_n}
                     C_\mn (x^a, x^i){\mid}_{x^i = 0} \
                      X^{k_1} ... \ X^{k_n}.
\label{taylorexp}
\end{eqnarray}
Constructing a Born-Infeld action adapted to the multiple brane case is a highly non-trivial problem. The Chern-Simons action however keeps a simple structure \cite{Tseytlin, Dorn, Hull}, besides showing several intricate properties.
One of these is the appearance of couplings proportional to a commutator of transverse scalars\cite{Myers}. These extra, purely non-Abelian couplings allow the brane to interact dielectrically with background fields of higher rank than the brane dimension. The multiple brane then expands into a higher-dimensional, fuzzy geometry. Such solutions are described in \cite{Tseytlin, Myers, TV}

The multiple D-brane action and its gauge properties were studied in\cite{vR, AJGL, AIJ,Adam}. An important observation was made in \cite{AIJ}: the matrix coordinates are affected by gauge transformations with the NS-NS parameter $\Sigma_1$:
\begin{eqnarray}
\delta_{\Sigma} V_a &=& - \Sigma_{\mu}DX^{\mu} \label{var_D0}\label{var_Sigma}\\
\delta_{\Sigma} X^{\mu} &=& i \Sigma_{\rho}[ X^{\rho}, X^{\mu}].\nonumber
\end{eqnarray}
This transformation is proportional to a commutator, such that it vanishes in the Abelian limit. A consequence is that every background field will undergo this non-Abelian NS-NS transformation as well, since the fields depend on the transverse coordinates.

In \cite{BJL, JL, JL2}, actions are established for the multiple fundamental string and the multiple gravitational wave. The following arguments suggest that these actions are valid in the strong coupling regime. Firstly, the large-$N$ limit ($N$ being the number of individual strings or waves forming the non-Abelian object) of the multiple wave matches the Abelian result for the giant graviton, thus suggesting that the multiple wave action describes a microscopical view of the giant graviton. A similar reasoning is true for the multiple fundamental string action. The large-$N$ limit thereof corresponds with the Abelian description of a dielectric D-brane. The second argument comes from the objects being BPS states, suggesting that no higher order corrections in $\alpha'$ are needed.

\begin{figure}
\begin{center}
\psfig{figure=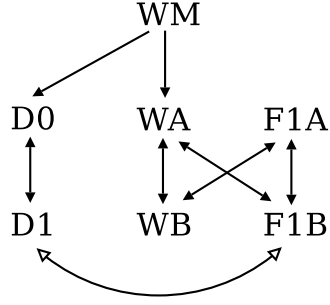, width=120pt} 
\caption{The various objects: the D-string (D1), the D0 brane (D0), the 11D wave (WM), the fundamental strings in IIA (F1A) and IIB (F1B); and the waves in IIA (WA) and IIB (WB). T-dualities are shown in double arrows. Single arrows denote dimensional reduction. The curved arrow represents S-duality.}
\label{dualiteiten}
\end{center}
\end{figure}

The actions of \cite{BJL, JL, JL2}, however, are only established to linear order in the background fields. The aim of this paper is to extend these actions to a fully gauge invariant form. The construction is made by duality relations and checked on gauge invariance. An overview of the dualisations is given in Figure \ref{dualiteiten}.

First we propose an action for the 11-D gravitational wave. This action is reduced along a direction other than the Taub-NUT direction of the wave, which results in a Type IIA wave (WA). T-duality yields a Type IIB fundamental string (FB) when performed along the isometry direction. Type IIB wave (WB) with two isometry directions results from a T-duality along an other direction. The Type IIA string (FA) can be reached from two directions: either by T-dualizing the IIB string, or by T-dualizing the IIB wave. The chain closes, providing a powerful check on the duality calculations. Notice that the duality chain leads to an effective S-duality transformation. Though the non-perturbative S-duality is not well defined in the case of multiple branes, the more rigorous chain here leads to the same result as a naive S-duality would.

Gauge transformations are well defined for the D-brane background fields. We will dualize the transformations as well as the fields to become corresponding gauge transformations for the exotic fields appearing in the multiple strings and waves. We will follow especially the transformations of the world-volume fields. The non-Abelian NS-NS transformation of the coordinate will dualize into a new transformation, following the transformations of the Born-Infeld vector and the exchange of the two-form fields $B_2$ and $C_2$.

The 11-dimensional wave is studied in section \ref{sect11D}. The next section concentrates on the IIA wave. The IIB and IIA string actions are derived in section \ref{sectFB} and \ref{sectFA}, respectively. The last action, the IIB wave, appears in section \ref{sectWB}. Appendices A-C list properties of the fields appearing in the various actions.

\sect{The eleven-dimensional multiple wave}\label{sect11D}

An action for 11D multiple waves is proposed in \cite{JL2}. Using gauge invariance, we extend this action to all orders in the background fields. The result is\footnote{Eleven-dimensional fields are denoted by a hatted letter}:

\begin{eqnarray}
\mathcal{L}_{WM} &=& \STr\Bigl\{ P \Bigl(\hat k^{-2} \hat k_1 \Bigr) - i P (\ii) \Bigl(\hat C_3 -3 \incl_k \hat C_3 \hat k^{-2}\hat k_1 \Bigr) \label{WMa}\\
& & - \frac{1}{2} P (\ii)^2 \Bigl( \incl_k \hat C_6 - 5\hat C_3 \incl_k \hat C_3 + 15 \incl_k \hat C_3 \incl_k \hat C_3 \hat k^{-2}\hat k_1 \Bigr) \nonumber \\
& & + \frac{i}{6} P (\ii)^3 \Bigl( \incl_k \hat C_8 -21 \incl_k \hat C_6 \incl_k \hat C_3 + 35 \hat C_3 \incl_k \hat C_3 \incl_k \hat C_3 - 105  \hat C_3 \incl_k \hat C_3 i_k \hat C_3 \hat k^{-2}\hat k_1 \Bigr) \nonumber  \\
& & + \frac{1}{24} P (\ii)^4 \Bigl( \incl_k \hat C_{10} - 36 \incl_k \hat C_8 \incl_k \hat C_3 + 378 \incl_k \hat C_6 \incl_k \hat C_3 \incl_k \hat C_3 - 315 \hat C_3 \incl_k \hat C_3 \incl_k \hat C_3 \incl_k \hat C_3 \nonumber \\ & & \quad\quad\quad\quad + 945  \incl_k \hat C_3 \incl_k \hat C_3 \incl_k \hat C_3 \hat k^{-2}\hat k_1 \Bigr) \Bigr\}. \nonumber
\end{eqnarray}

$\hat k$ denotes a Killing vector which represents the Taub-NUT direction:
\begin{eqnarray}
\hat k^{\mu} = \delta^{\mu}\!_{11} &,\quad& \hat k_{\mu} =  \hat g_{\mu 11}.
\end{eqnarray}
An inclusion of a form with the Killing vector extracts the components which have one index equal to $11$. The notation is the same as for the inclusions with transverse scalars:
\begin{eqnarray}
\incl_k \hat C_p &=& \hat k^{\rho}\hat C_{\rho \mu_2 ... \mu_{p}} = \hat C_{11, \mu_2...\mu_p}.
\end{eqnarray}

In addition to being charged under a Killing vector field, the multiple wave couples dielectrically to forms of higher rank. This dielectric coupling is proportional to commutators and characteristic for multiple branes, strings and waves. While the higher-rank forms featuring in the Myers action are the Ramond-Ramond potentials, more exotic eleven-dimensional forms appear here. $\hat C_8$ and $\hat C_{10}$ are the fields that minimally couple, respectively, to the M-theory Kaluza-Klein monopole and the M9 brane (see e.g. \cite{EL} for a discussion of these forms). A list of all exotic forms and the respective branes is included as appendices A-C.

Because of the appearance of the Killing vector, the sigma-model is gauged with respect to translation symmetry of the eleventh coordinate. In \cite{BJL, JL, JL2}, covariant pull-backs were used to get a more Lorentz-invariant look. To non-linear order, however, the Killing vector appears not only pull-backed but also contracted with a commutator. Defining modified commutators is possible, but it is easier and more transparent to see the Killing vector as a background field.

A more compact way of writing (\ref{WMa}) uses the property $\incl_k( \hat k^{-2}\hat k_1) = \frac{g_{11,11}}{g_{11,11}} = 1$, such that for every $p$-form $T_p$:
\begin{eqnarray}
(p+1) \incl_k (\hat T_p \hat k^{-2}\hat k_1) &=& p\ \incl_k \hat T_p\ \hat k^{-2}\hat k_1 + (-)^{p} \hat T_p \label{compact}.
\end{eqnarray}

This proves to be useful, especially when dealing with multiple isometry directions, such as in the IIB multiple wave.
Written in this way, the action (\ref{WMa}) becomes:

\begin{eqnarray}
\mathcal{L}_{WM} &=& \STr\Bigl\{ P \Bigl( \hat k^{-2} \hat k_1 \Bigr) + i P (\ii) \incl_k \Bigl(4\hat C_3 \hat k^{-2}\hat k_1 \Bigr) \label{WM}\\
& & + \frac{1}{2} P (\ii)^2  \incl_k \Bigl[ 6 \Bigl(\incl_k \hat C_6 - 5\hat C_3 \incl_k \hat C_3 \Bigr) \hat k^{-2}\hat k_1 \Bigr] \nonumber \\
& & - \frac{i}{6} P (\ii)^3 \incl_k \Bigl[ 8 \Bigl(\incl_k \hat C_8 -21 \incl_k \hat C_6 \incl_k \hat C_3 + 35 \hat C_3 \incl_k \hat C_3 \incl_k \hat C_3 \Bigr) \hat k^{-2}\hat k_1 \Bigr] \nonumber  \\
& & - \frac{1}{24} P (\ii)^4 \incl_k \Bigl[ 10 \Bigl( \incl_k \hat C_{10} - 36 \incl_k \hat C_8 \incl_k \hat C_3 + 378 \incl_k \hat C_6 \incl_k \hat C_3 \incl_k \hat C_3 - 315 \hat C_3 \incl_k \hat C_3 \incl_k \hat C_3 \incl_k \hat C_3 \Bigr) \hat k^{-2}\hat k_1 \Bigr] \Bigr\} \nonumber.
\end{eqnarray}
The invariance of the action under coordinate transformations in the isometry direction is now manifest, since everything is written as an inclusion with $\hat k_1$.

A remarkable symmetry of the action is under the gauge transformation of the field $\incl_k \hat C_3$. The Born-Infeld vector and the transverse scalar transform respectively with a shift and a commutator:
\begin{eqnarray}
\delta \hat V_a &=&  (\incl_k \hat \Lambda_2)_{\rho}D_a X^{\rho} \label{var_W11}\\
\delta \hat X^{\mu} &=& -i(\incl_k \hat \Lambda_2)_{\rho}\lbrack X^{\rho}, X^{\mu}\rbrack.\nonumber
\end{eqnarray}

A dimensional reduction of the action (\ref{WM}) along the Killing direction results in the well-known Myers action for multiple D0-branes. Hereby the Killing vector takes the role of the Kaluza-Klein vector $C_1$. The gauge transformations (\ref{var_W11}) reduce to the NS-NS variations (\ref{var_Sigma}) for the worldvolume fields of the multiple D0-brane. So the Myers action can be seen as a weak coupling limit of the (\ref{WM}) action, thereby motivating the interpretation of the latter as an action for multiple gravitational waves in M-theory.

\sect{The IIA multiple wave}\label{sectWA}

When reducing the action (\ref{WM}) along a direction other than the Taub-NUT direction, one finds the Chern-Simons action for the IIA multiple wave. This was carried out to linear order in \cite{JL}. Using the full reduction rules we get the following result. 

We assume that the reduced dimension is orthogonal to the Taub-NUT direction of the wave. This restriction is necessary to close the duality chain. Closure requires the T-dualities to commute, which is the case for orthogonal directions. The orthogonality is assured by setting $\incl_z (\hat k^{-2}\hat k_1) = \frac{g_{z 11}}{g_{11,11}}$ to zero. $\hat k^{-2}\hat k_1$ reduces then to $k^{-2}k_1 = \frac{ g_{\mu 11}}{ g_{11,11}}$. 

The reduced coordinate $X^z$ becomes a scalar which we call $S$. So
\begin{eqnarray}
X^z \rightarrow S &\quad\quad& DX^z \rightarrow DS.
\end{eqnarray}

With these assumptions the Type IIA wave action becomes:
\begin{eqnarray}
\mathcal{L}_{WA} &=& \STr\Bigl\{  P\Bigl( k^{-2} k_1 \Bigr) - i \left( DS (\ii) - P \incl_{\lbrack S,X\rbrack}\right) \incl_k \Bigl( 3 B_2  k^{-2} k_1 \Bigr) \label{WA} \\
& & + i P (\ii) \incl_k \Bigl( 4 C_3 k^{-2} k_1 \Bigr) + \frac{1}{2}\left( DS (\ii)^2 - 2 P(\ii)\incl_{\lbrack S,X \rbrack} \right) \incl_k\Bigl[ 5(i_k C_5 + 4 C_3 i_k B_2) k^{-2} k_1\Bigr] \nonumber \\  
& & + \frac{1}{2} P (\ii)^2  \incl_k\Bigl[ 6 \Bigl(\incl_k  B_6 - 5  C_3 \incl_k  C_3 \Bigr)  k^{-2} k_1\Bigr] \nonumber \\ 
& & - \frac{i}{6} \left( DS (\ii)^3 - 3 P (\ii)^2\incl_{\lbrack S, X \rbrack}\right) \incl_k\Bigl[7 \Bigl(\incl_k  N_7 + 6\incl_k  B_6 \incl_k  B_2 - 15 \incl_k  C_5 \incl_k  C_3 \nonumber \\ & & \quad \quad \quad \quad- 30  C_3 \incl_k  C_3 \incl_k  B_2 \Bigr)  k^{-2} k_1\Bigr] \nonumber  \\
& & - \frac{i}{6} P (\ii)^3 \incl_k\Bigl[ 8 \Bigl(\incl_k  N_8 - 21 \incl_k  B_6 \incl_k  C_3 + 35  C_3 (\incl_k  C_3)^2 \Bigr) k^{-2} k_1\Bigr] \nonumber \\ 
& & - \frac{1}{24}\left( DS (\ii)^4 - 4 P (\ii)^3i_{\lbrack S, X \rbrack} \right) i_k \Bigl[ 9 \Bigl( \incl_k  N_9 + 8 \incl_k  N_8 \incl_k  B_2 - 28 \incl_k  N_7 \incl_k  C_3 \nonumber \\ & & \quad \quad \quad \quad - 168 \incl_k  B_6 \incl_k  C_3 \incl_k  B_2  + 210 \incl_k  C_5 (\incl_k  C_3)^2  + 280  C_3 (\incl_k  C_3)^2 \incl_k  B_2 \Bigr) k^{-2} k_1\Bigr] \nonumber \\
& & - \frac{1}{24} P (\ii)^4  \incl_k \Bigl[ 10 \Bigl(\incl_k  N_{10} - 36 \incl_k  N_8 \incl_k  C_3 + 378 \incl_k  B_6 (\incl_k  C_3)^2 - 315  C_3 (\incl_k  C_3)^3 \Bigr)  k^{-2} k_1\Bigr] \Bigr\} \nonumber
\end{eqnarray}

The wave couples minimally to the Killing vector and dielectrically to higher-order forms. Gravitational and exotic background fields appear in these dielectric couplings. We will encounter yet more exotic fields in the IIB string and IIB wave actions discussed in the following sections. Their properties are known from duality relations and are listed in the appendix.

Gauge invariance has been checked. Since $\incl_k \hat \Lambda_2$ reduces to both $\incl_k  \Lambda_2$ and $\incl_k \Sigma_1$, we will encounter these two parameters in the variation of the Born-Infeld vector, the transverse scalars and the scalar $S$. 

The worldvolume field variations are:
\begin{eqnarray}
\delta V_a &=& D_a X^{\rho}(\incl_k  \Lambda_2)_{\rho} + D_a S \incl_k  \Sigma_1\label{var_WA}\\
\delta X^{\mu} &=&-i (\incl_k \Lambda_2)_{\rho}[X^{\rho}, X^{\mu}] -i \incl_k  \Sigma_{1}[S, X^{\mu}]\nonumber\\
\delta S &=& - \Lambda_0 - i (\incl_k  \Lambda_{2})_{\rho}[X^{\rho}, S]. \nonumber
\end{eqnarray}


\sect{The multiple $F1$ in Type IIB}\label{sectFB}

When T-dualized along its Taub-NUT direction, the IIA gravitational wave becomes a fundamental string in Type IIB. Hereby the worldvolume fields are mapped as usual. The scalar $S$ becomes a component of the Born-Infeld vector:
\begin{eqnarray}
S &\longleftrightarrow & -V_x.
\end{eqnarray}

The duality was carried out to linear order in \cite{BJL}. Again, using the reduction and uplifting rules on the action (\ref{WA}) we recover a fully gauge invariant action of the multiple fundamental string in Type IIB. 

\begin{eqnarray}
\mathcal{L}_{FB} &=& \STr \Bigl\{ P( B_2 ) + i (\ii) ( B_2 ) \wedge  F + i P (\ii) ( C_4 ) - \frac{1}{2} (\ii)^2 ( C_4 )\wedge  F \label{FB} \\
& & + \frac{1}{2} P (\ii)^2 ( B_6 ) + \frac{i}{6} (\ii)^3( B_6 ) \wedge  F - \frac{i}{6} P (\ii)^3 ( Q_8 ) \nonumber \\
& &  + \frac{1}{24} (\ii)^4 ( Q_8)\wedge  F - \frac{1}{24} P (\ii)^4 ( Q_{10} ) \Bigr\} \nonumber
\end{eqnarray}

Notice that there appear no isometry directions any more. The multiple fundamental string in IIB has the full ten-dimensional Lorentz-invariance, just like a single fundamental string.

As for the variation parameters, T-duality maps both $ \Lambda_0$ and $\incl_k  \Lambda_2$ onto $\incl_k \Lambda_1$, following the recombination of the IIA fields $V_a$ and $S$ into the IIB Born-Infeld vector $V_a$. The $\incl_k \Sigma_1$ variation becomes a coordinate transformation. So the gauge variation of the worldvolume fields goes only with $ \Lambda_1$.
\begin{eqnarray}
\delta V_a &=& - \Lambda_{\rho}D_a X^{\rho}\label{var_FB}\\
\delta X^{\mu} &=& i \Lambda_{\rho}[X^{\rho}, X^{\mu}] \nonumber
\end{eqnarray}

Now it is very clear how the twoforms are interchanged with respect to the D-string. The worldvolume fields vary with the R-R parameter. As was already mentioned in \cite{BJL}, the Born-Infeld vector of the D-branes is changed by S-duality into another worldvolume vector. In the Abelian limit the field strength of this vector combines with the pull-backed R-R twoform to an invariant field strength $\mathcal{F}' = 2\partial V' + P[C_2]$. 


\sect{Transverse T-duality: a Type IIA fundamental string}\label{sectFA}
As mentioned in \cite{BJL}, a transverse T-dualization of the Type IIB fundamental string gives a fundamental string with winding number in Type IIA.  

\begin{eqnarray}
\mathcal{L}_{FA} &=& \STr\lbrace \left(-2D\omega P  + i  F \incl_{[ \omega , X]} \right) \Bigl( l^{-2}l_1 \Bigr) + \left( P + i F (\ii)\right) \incl_l \Bigl(3 B_2 l^{-2}l_1 \Bigr) \label{FA} \\
& & + i\left( -2 D\omega P (\ii) + P \incl_{\lbrack \omega,X\rbrack} + i F (\ii)\incl_{[ \omega, X ]} \right) \incl_l \Bigl(4 C_3  l^{-2}l_1 \Bigr) \nonumber \\
& & + i \left( P (\ii) +\frac{i}{2} F (\ii)^2\right) \incl_l \Bigl[ 5\Bigl(\incl_l  C_5 + 4 C_3 \incl_l  B_2 \Bigr) l^{-2}l_1\Bigr] \nonumber \\ 
& & +\frac{1}{2}\left(-2 D\omega P (\ii)^2 +2 P(\ii)i_{[ \omega,X ]} +i  F (\ii)^2 \incl_{[ \omega, X ]}\right) \incl_l\Bigl[ 6\Bigl(\incl_l  B_6 - 5  C_3 \incl_l  C_3 \Bigr)  l^{-2}l_1\Bigr] \nonumber \\
& & + \frac{1}{2} \left( P(\ii)^2 +\frac{i}{3} F (\ii)^3\right) \incl_l\Bigl[ 7\Bigl(\incl_l  N_7 + 6\incl_l  B_6 \incl_l  B_2 - 15 \incl_l  C_5 \incl_l  C_3 - 30  C_3 \incl_l  C_3 \incl_l  B_2 \Bigr) l^{-2}l_1\Bigr]\nonumber \\
& & -\frac{i}{6}\left( -2D\omega P(\ii)^3 +3 P (\ii)^2 \incl_{[ \omega,X]} +iF (\ii)^3 \incl_{[ \omega,X ]} \right)\nonumber \\ & & \quad \quad \incl_l \Bigl[ 8 \Bigl(\incl_l  N_8 - 21 \incl_l  B_6 \incl_l  C_3 + 35  C_3 (\incl_l  C_3)^2\Bigr)  l^{-2}l_1\Bigr] \nonumber \\
& & - \frac{i}{6} P (\ii)^3 \incl_l \Bigl[ 9 \Bigl(\incl_l  N_9 + 8 \incl_l  N_8 \incl_l  B_2 - 28 \incl_l  N_7 \incl_l  C_3 - 168 \incl_l  B_6 i_l  C_3 \incl_l  B_2 + 210 \incl_l  C_5 (\incl_l  C_3)^2 \nonumber \\ & & \quad \quad\quad\quad + 280  C_3 (\incl_l  C_3)^2 \incl_l  B_2 \Bigr)  l^{-2}l_1\Bigr] \nonumber\\
& & - \frac{1}{6}P (\ii)^3 i_{[ \omega,X ]}\incl_l \Bigl[ 10\Bigl(\incl_l  N_{10} - 36 \incl_l  N_8 \incl_l  C_3 + 378 \incl_l  B_6 (\incl_l  C_3)^2 - 315  C_3 (\incl_l  C_3)^3 \Bigr) l^{-2}l_1\Bigr] \Bigr\}. \nonumber
\end{eqnarray}

The T-duality rules are the same as the ones used in Section \ref{sectFB}. Indeed, both dualities connect a IIB theory to the IIA with one isometry direction. To make clear that the isometry directions fulfill a different role in the IIA string and the IIA wave, we denote their Killing vectors with different symbols. So $k$ is the IIA wave's Taub-NUT direction, while $l$ denotes the string's wrapping direction. In both cases the isometry directions allow for exotic form fields.

The worldvolume gauge variations are now
\begin{eqnarray}
\delta V_a &=&  (i_l  \Lambda_2)_{\rho}D_a X^{\rho} +  \Lambda_0 D_a \omega\label{var_FA}\\
\delta X^{\mu} &=& -i (i_l  \Lambda_2)_{\rho} [X^{\rho}, X^{\mu}] - i  \Lambda_0 [\omega, X^{\mu}] \nonumber \\
\delta \omega &= & -\incl_l \Sigma_1 - i (\incl_l \Lambda_2)_{\rho}[X^{\rho}, \omega]. \nonumber
\end{eqnarray}


\sect{The Type IIB gravitational wave}\label{sectWB}
The IIB gravitational wave can be reached by a transverse T-duality on the IIA wave, but also via a T-duality along a worldvolume direction of the IIA fundamental string. This two ways should give the same result, closing the chain of dualities. This is indeed the case, up to a sign difference of the R-R fields. This sign difference comes from interchanging the two T-duality directions, as mentioned in \cite{JL}. The action mentioned here is the one coming from the transverse T-duality on the IIA wave, in order to compare with \cite{JL}. In addition to the duality rules we used for the IIB string, we need now more dualities connecting to the exotic IIB fields with two isometries. Due to our assumption of orthogonal isometry directions, we have that $\incl_l (k^{-2}k_1) = \incl_k (l^{-2}l_1) = \incl_k\incl_l B_2 = 0$. This taken into account, the IIB wave action is:

\begin{eqnarray}
\mathcal{L}_{WB} &=& \STr \Bigl\{ i \left( D\omega \incl_{[ S,X]} - DS \incl_{[ \omega,X]} - [ S, \omega] P \right) ( l^{-2}l_1) + P( k^{-2}k_1) \label{WIIB} \\ 
& & + i \left( DS(\ii) -P \incl_{[S,X]}\right) \incl_k \incl_l \Bigl(12 B_2  k^{-2}k_1  l^{-2}l_1\Bigr)\nonumber \\ & &  + i\left( D\omega(\ii) -P \incl_{[ \omega,X]}\right) \incl_k \incl_l \Bigl(12 C_2  k^{-2}k_1  l^{-2}l_1\Bigr) \nonumber \\
& & + \left( D\omega (\ii) \incl_{[ S,X]} - DS (\ii) \incl_{[\omega,X]} + P \incl_{[ S,X]} \incl_{[ \omega ,X ]} - [S, \omega] P(\ii) \right)\nonumber \\ && \quad \quad \incl_k \incl_l \Bigl[ 20\Bigl(\incl_k  C_4 - 3  C_2 \incl_k  B_2\Bigr) k^{-2}k_1  l^{-2}l_1 \Bigr] \nonumber\\ 
& & + i P (\ii) \incl_k \incl_l \Bigl[ 20\Bigl(\incl_l  C_4 - 3  C_2 \incl_l  B_2\Bigr) k^{-2}k_1  l^{-2}l_1 \Bigr] \nonumber \\ 
& & -\frac{1}{2}\left( DS(\ii)^2- 2P (\ii)\incl_{[S,X]}\right) \incl_k \incl_l \Bigl[ 30\Bigl(\incl_k \incl_l C_6 + 4 \incl_l C_4 \incl_k B_2 - 4 \incl_k  C_4 \incl_l  B_2 \nonumber \\ & & \quad \quad \quad \quad - 12 C_2 \incl_l  B_2 \incl_k  B_2\Bigr)  k^{-2}k_1  l^{-2}l_1 \Bigr] \nonumber \\
& & - \frac{1}{2} \left( D\omega(\ii)^2 - 2 P (\ii)\incl_{[\omega,X]}\right) \incl_k \incl_l \Bigl[ 30\bigl( \incl_k \incl_l  B_6 + 6 \incl_k \incl_l  C_4  C_2\Bigr)  k^{-2}k_1  l^{-2}l_1 \Bigr] \nonumber \\
& & - \frac{i}{2}\left( D\omega (\ii)^2 \incl_{[S,X]} - DS (\ii)^2 \incl_{[ \omega,X]} + 2 P (\ii) \incl_{[S,X]} \incl_{[\omega ,X ]} - [S, \omega] P(\ii)^2 \right) \nonumber \\
& & \quad \quad \incl_k \incl_l \Bigl[ 42\bigl(\incl_k \incl_l  N_7 + 5 \incl_k \incl_l  B_6 \incl_k  B_2 - 5 \incl_k  C_4 \incl_k \incl_l  C_4 + 30 \incl_k \incl_l  C_4  C_2 \incl_k  B_2\Bigr)  k^{-2}k_1  l^{-2}l_1 \Bigr] \nonumber \\
& & + \frac{1}{2}P (\ii)^2 \incl_k \incl_l \Bigl[ 42\Bigl(\incl_k \incl_l  \mathcal{N}_7 - 5 \incl_l  C_4 \incl_k \incl_l  C_4 + 5 \incl_k \incl_l  B_6 \incl_l  B_2 + 30 \incl_k \incl_l  C_4  C_2 \incl_l  B_2\Bigr) k^{-2}k_1  l^{-2}l_1 \Bigr] \nonumber \\ 
& & + \frac{i}{6} \left( DS(\ii)^3 - 3P (\ii)^2 \incl_{[S,X]}\right) \incl_k \incl_l \Bigl[ 56\Bigl(\incl_k \incl_l  \mathcal{N}_8 + 6 \incl_k \incl_l  \mathcal{N}_7\incl_k  B_2 - 6 \incl_k \incl_l  N_7 \incl_l  B_2 \nonumber \\ & & \quad\quad\quad\quad - 30 \incl_k \incl_l  B_6 \incl_k  B_2 \incl_l  B_2 - 180 \incl_k \incl_l  C_4  C_2 \incl_k  B_2 \incl_l  B_2 - 15 \incl_k \incl_l  C_4 \incl_k \incl_l  C_6 \nonumber \\ & & \quad\quad\quad\quad - 30 \incl_l  C_4 \incl_k \incl_l  C_4 \incl_k  B_2 + 30 \incl_k  C_4 \incl_k \incl_l  C_4 \incl_l  B_2\Bigr) k^{-2}k_1  l^{-2}l_1 \Bigr] \nonumber \\
& & + \frac{i}{6}\left( D\omega(\ii)^3 - 3 P (\ii)^2\incl_{[\omega,X]}\right) \incl_k \incl_l \Bigl[ 56\Bigl(\incl_k \incl_l  N_8 - 45  C_2 \incl_k \incl_l  C_4 \incl_k \incl_l  C_4\Bigr) k^{-2}k_1  l^{-2}l_1 \Bigr] \nonumber \\
& & - \frac{1}{6}\left( D\omega (\ii)^3 \incl_{[S,X]} - DS (\ii)^3 \incl_{[ \omega,X ]} + 3 P (\ii)^2 \incl_{[S,X ]} \incl_{[\omega ,X ]} - [S, \omega] P(\ii)^3 \right)\nonumber \\
& & \quad \quad \incl_k \incl_l \Bigl[72\Bigl( \incl_k \incl_l N_9 + 7 \incl_k \incl_l  N_8 \incl_k  B_2 - 21 \incl_k \incl_l  N_7 \incl_k \incl_l  C_4 \nonumber \\ & & \quad\quad\quad\quad + 35 \incl_k  C_4 (\incl_k \incl_l  C_4)^2 - 315  C_2 (\incl_k \incl_l  C_4)^2 \incl_k  B_2\Bigr) k^{-2}k_1  l^{-2}l_1 \Bigr] \nonumber \\
& & - \frac{i}{6} P (\ii)^3 \incl_k \incl_l \Bigl[ 72\Bigl( \incl_k \incl_l  \mathcal{N}_9 + 7 \incl_k \incl_l  N_8 \incl_l  B_2 - 21 \incl_k \incl_l  \mathcal{N}_7 \incl_k \incl_l  C_4 + 35 \incl_l  C_4 (\incl_k \incl_l  C_4)^2 \nonumber \\ & & \quad\quad\quad\quad - 315  C_2 (\incl_k \incl_l  C_4)^2 \incl_l  B_2\Bigr) k^{-2}k_1  l^{-2}l_1 \Bigr] \nonumber \\
& & - \frac{1}{6} P (\ii)^3 \incl_{[S,X]} \incl_k \incl_l \Bigl[90\Bigl( \incl_k \incl_l  \mathcal{N}_{10} - 8 \incl_k \incl_l N_9 \incl_l  B_2 + 8 \incl_k \incl_l \mathcal{N}_9 \incl_k  B_2 - 56 \incl_k \incl_l  N_8 \incl_k  B_2 \incl_l  B_2 \nonumber \\ & & \quad\quad\quad\quad - 28 \incl_k \incl_l  \mathcal{N}_8 \incl_k \incl_l  C_4 - 168 \incl_k \incl_l  \mathcal{N}_7 \incl_k \incl_l  C_4 \incl_k  B_2 + 168 \incl_k \incl_l  N_7 \incl_k \incl_l  C_4 \incl_l  B_2 \nonumber \\ & & \quad\quad\quad\quad + 210 \incl_k \incl_l C_6 (\incl_k \incl_l  C_4)^2 - 280 \incl_k  C_4 (\incl_k \incl_l  C_4)^2  \incl_l  B_2 + 280 \incl_l  C_4 (\incl_k \incl_l  C_4)^2 \incl_k  B_2 \nonumber \\ & & \quad\quad\quad\quad+ 2520  C_2 (\incl_k \incl_l  C_4)^2 \incl_k  B_2 \incl_l  B_2\Bigr) k^{-2}k_1  l^{-2}l_1 \Bigr]\nonumber \\
& & -\frac{1}{6} P (\ii)^3 \incl_{[\omega,X]} \incl_k \incl_l \Bigl[ 90\Bigl( \incl_k \incl_l  N_{10} + 420  C_2 (\incl_k \incl_l  C_4)^3 \Bigr) k^{-2}k_1  l^{-2}l_1 \Bigr] \Bigr\} .\nonumber
\end{eqnarray}

Not only the action but also the transformations get a more complicated form due to the two isometries. T-duality of the IIA transformations learns us that three parameters affect the Born-Infeld field and the transverse scalars, while the scalars $\omega$ and $S$ shift with two more parameters.
\begin{eqnarray}
\delta V_a &=& -(\incl_k \incl_l  \Lambda_3)_{\rho }D_a X^{\rho} + (\incl_k  \Lambda_1)D_a \omega + (\incl_k  \Sigma_1) D_a S \label{var_WB}\\
\delta X^{\mu} &=& i (\incl_k \incl_l  \Lambda_3)_{\rho} [X^{\rho}, X^{\mu}]-i (\incl_k  \Lambda_1)[\omega, X^{\mu}] -i (\incl_k \Sigma_1)[S, X^{\mu}]\nonumber\\
\delta \omega &=& - \incl_l  \Sigma_1 -i (\incl_k \incl_l  \Lambda_3)_{\rho} [X^{\rho}, \omega] - (\incl_k \Sigma_1)[S, \omega]\nonumber\\
\delta S &=& \incl_l  \Lambda_1-i (\incl_k \incl_l  \Lambda_3)_{\rho} [X^{\rho},S] - (\incl_k \Lambda_1)[\omega, S]\nonumber
\end{eqnarray}
The most remarkable feature of this action and its variation is their complexity, especially when compared to the IIB fundamental string. The reason of this lies indeed in the isometries. In fact, the IIB wave action is eight-dimensional instead of ten-dimensional. The different role played by the parameters with an $\incl_k$ inclusion and an $\incl_l$ inclusion points at the different nature of the isometries. $k$ is the Taub-NUT direction while $l$ is the wrapping direction. $\incl_k \incl_l  \Lambda_3$ is the wave's counterpart of $ \Lambda_1$. All other variations appear only due to the isometries. Indeed, they are always proportional to the scalars $S$ and $\omega$.

\sect{Discussion}\label{sectdisc}
In this paper we have discussed fully gauge invariant actions for the multiple 11D gravitational wave, the Type IIA and IIB wave and the IIA and IIB string. Special attention is drawn to the world volume fields and their non-Abelian gauge variations. 

Another issue, not yet mentioned, is the role a mass parameter would take. Multiple D-branes in massive Type IIA \cite{Romans} have been studied in \cite{AJGL}. Two ways can be followed to extend the Type IIA actions here with a mass: either perform a massive T-duality on the Type IIB string, or start with the D0-brane in massive Type IIA and following the chain used here again. The latter method involves BLO theory \cite{BLO}. Both methods would yield a gravitational wave in Romans theory, where $\incl_k C_3$ takes the role of the massive field.

\newpage
\vspace{1cm}

\noindent
{\bf Acknowledgments}\\
We wish to thank Walter Troost for the useful discussions. Bert Janssen and Yolanda Lozano merit a special thank for their help in understanding and interpreting the formulas.
This work is supported in part by the European Community's Human
Potential Programme under contract MRTN-CT-2004-005104 `Constituents,
fundamental forces and symmetries of the universe'. J.A. is Aspirant FWO Vlaanderen.
Her work is supported in part by the FWO - Vlaanderen, project
G.0235.05 and by the Federal Office for Scientific, Technical and
Cultural Affairs through the "Interuniversity Attraction Poles Programme
-- Belgian Science Policy" P5/27.

\appendix

\renewcommand{\theequation}{\Alph{section}.\arabic{equation}}

\sect{Exotic forms}
In the following tables we will list the background fields together with the branes which minimally couple to them. Most of the exotic branes are studied in \cite{MeessenOrtin} and \cite{Hull2}. In the latter paper a coding system was developed to characterize the branes. We will list both the name (for the branes found in \cite{MeessenOrtin}) and the code, in addition to the corresponding background field.

In M theory:
\begin{equation}
\begin{array}{c|c|c}
\mbox{Name}&\mbox{Code}&\mbox{Background field}\\
\hline
WM &(1^{-1}) & \hat k^{-1}\hat k_2\\
M2 &(2) & \hat C_3\\
M5 & (5) & \hat C_6\\
KK7M &(6,1^2)  & \incl_k\hat C_8\\
KK9M & (8,1^3) & \incl_k \hat C_{10}
\end{array}
\end{equation}

In type IIA:
\begin{equation}
\begin{array}{c|c|c}
\mbox{Name}&\mbox{Code}&\mbox{Background field}\\
\hline
Dp& (p;1)& C_{p+1}\\
WA &(1^{-1};0) & k^{-1}k_2\\
F1A &(1;0) & B_2\\
NS5A & (5;2) & B_6\\
KK6A &(5,1^2;2) &\incl_k N_7\\ 
KK7A & (6,1^2;3) & \incl_k N_8\\
KK8A & (7,1^3;3) & \incl_k N_9\\
KK9A & (8,1^3;4) & \incl_k N_{10}
\end{array}
\end{equation}

In type IIB:
\begin{equation}
\begin{array}{c|c|c}
\mbox{Name}&\mbox{Code}&\mbox{Background field}\\
\hline
Dp & (p;1)& C_{p+1}\\
WB & (1^{-1};0) & k^{-1}k_2\\
F1B &(1;0) & B_2\\
NS5B & (5;2) & B_6\\
KK6B &(5,1^2;2) &\incl_k N_7 / \incl_l \mathcal{N}_7\\
Q7 & (7;3)& Q_8\\
 & (5,2^2;2) & \incl_k\incl_l\mathcal{N}_8\\
 & (5,2^2;3) & \incl_k \incl_lN_8\\
KK8B & (6,1^2,1^3;3) & \incl_k\incl_lN_9 / \incl_k\incl_l\mathcal{N}_9\\
Q9 & (9;4) & Q_{10}\\
 & (7,2^3;3) & \incl_k\incl_l \mathcal{N}_{10}\\
 &(7,2^3;4) & \incl_k\incl_l N_{10}
\end{array}
\end{equation}

\begin{figure}
\begin{center}
\psfig{figure=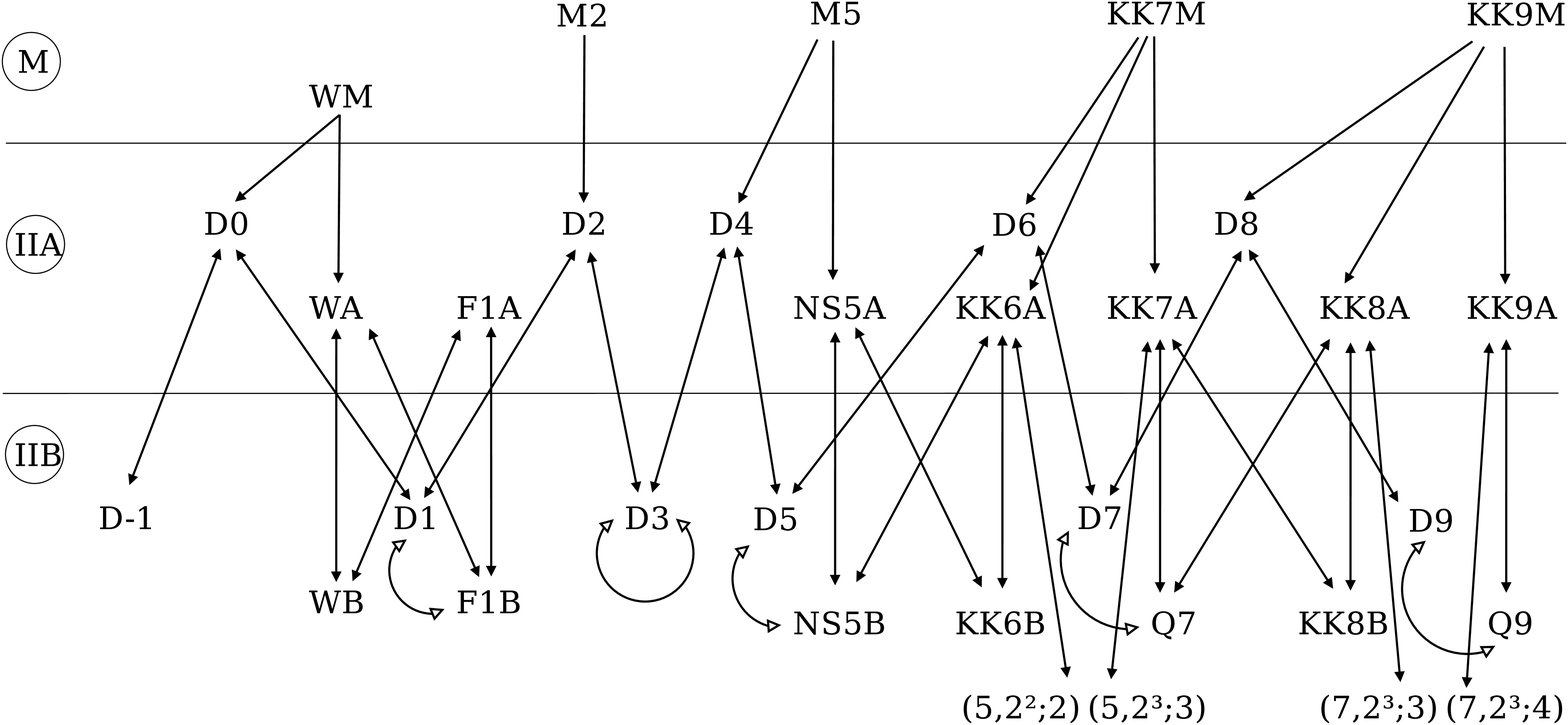, width=500pt, angle = 90}
\caption{Vertical single arrows denote direct dimensional reduction, diagonal single arrows stand for double dimensional reduction. Double straight arrows represent T-duality. Double curved arrows denote S-duality.}\label{exoticbranes}
\end{center}
\end{figure}

For the sake of clarity, we will include a diagram depicting the duality relations between the branes (see Fig. \ref{exoticbranes}).

\sect{Reduction and T-duality rules}\label{sectred}
Here we list the reduction and T-duality rules we used. They can also be found, up to linear order, in \cite{JL, JL2}. Most of them to are mentioned to higher order in \cite{EL}. Some sign conventions may differ.

Reduction rules for the eleven-dimensional fields along the isometry direction are:
\begin{equation}
\begin{array}{rclcrcl} 
\hat k^{-2} \hat k_1 &=&  C_1 & & & & \\
\incl_k \hat C_3 &=&  B_2 &\quad& \hat C_3 &=&  C_3\\
\incl_k \hat C_6 &=&  C_5 - 5  C_3  B_2 &\quad& & & \\
\incl_k \hat C_8 &=&  C_7 - 35  C_3  B_2  B_2 & & & & \\
\incl_k \hat C_{10} &=&  C_9 - 315  C_3  B_2  B_2  B_2. &\quad& & & 
\end{array}
\label{red11D-k}
\end{equation}

These are the reduction rules for the eleven-dimensional fields along a direction other than the isometry direction. The reduced-over direction is denoted by the letter $z$. In analogy to the inclusion $\incl_k$ we defined the inclusion $\incl_z$ as being $\incl_z \hat C_p = C_{z \mu_2 ... \mu_p}$.

\begin{equation}
\begin{array}{rclcrcl} 
\incl_z (\hat k^{-2}) \hat k_1 &=& 0 &\ &\hat k^{-2} \hat k_1 &=& k^{-2} k_1 \\
\incl_k \incl_z \hat C_3 &=&  \incl_k  B_2 &\ & \incl_k \hat C_3 &=& \incl_k  C_3\\
\incl_z \hat C_3 &=&  B_2 &\ & \hat C_3 &=&  C_3 \\
\incl_z \hat C_6 &=& C_5 -5 C_3 B_2 &\ & \hat C_6 &=& B_6\\
\incl_k \incl_z \hat C_8 &=& \incl_k  N_7 - 30 (\incl_k C_3)^2 B_2 +20 \incl_k C_3 C_3 \incl_k  B_2 &\ & \incl_k \hat C_8 &=& \incl_k N_8\\
\incl_k \incl_z \hat C_{10} &=&  \incl_k  N_9 - 315 (\incl_k  C_3)^3 B_2 + 210 C_3 (\incl_k  C_3)^2 \incl_k B_2 &\ & \incl_k \hat C_{10}&=& \incl_k  N_{10}.
\end{array}
\label{red11D-z}
\end{equation}

The following list contains the T-duality rules between the IIA fields with $k$ as isometry direction, on the right-hand side, and the IIB potentials with no isometry on the left-hand side. 
\begin{eqnarray}
\frac{g_z}{g_{zz}} &=& -\incl_k  B_2\label{T-IIA-k} \\
- \incl_z  B_2 &=& k^{-2}k_1 \nonumber\\
 B_2 &=&  B_2 + 2 \incl_k  B_2 k^{-2}k_1\nonumber\\
\incl_z  C_{2n} &=&  C_{2n-1} -(2n-1) \incl_k C_{2n-1}k^{-2}k_1\nonumber\\
C_{2n} &=& \incl_k C_{2n+1} + 2n( C_{2n-1} - (2n-1) \incl_k C_{2n} k^{-2}k_1)\incl_k B_2\nonumber\\
\incl_z  B_6 &=& \incl_k B_6 - 5 ( C_3 -3\incl_k C_3 k^{-2}k_1) \incl_k C_3 \nonumber\\
B_6 &=& \incl_k N_7 + 6(\incl_k B_6 - 5 (C_3-3\incl_k C_3k^{-2}k_1)\incl_k C_3) \incl_k B_2 - 15 \incl_k C_5 \incl_k C_3 \nonumber \\
\incl_z Q_8 &=& \incl_k N_8 - 21 \incl_k B_6\incl_k C_3 +35 (C_3 - 3\incl_k C_3 k^{-2}k_1)(\incl_k C_3)^2 \nonumber \\
Q_8 &=& \incl_k N_9 + 8 (\incl_k N_8 - 21 \incl_k B_6\incl_k C_3 +35 (C_3 - 3\incl_k C_3 k^{-2}k_1)(\incl_k C_3)^2)\incl_k B_2 \nonumber \\ & &  - 27 \incl_k N_7 \incl_k C_3 + 210 \incl_k C_5 (\incl_k C_3)^2 \nonumber \\
i_z Q_{10} &=& i_k N_{10} - 36 \incl_k N_8 \incl_k C_3 + 378 \incl_k B_6 (\incl_k C_3)^2 -315 (C_3 -\incl_k C_3 k^{-2}k_1)(\incl_k C_3)^3\nonumber
\end{eqnarray}

Here we list the T-duality rules between the exotic Type IIB potentials with two isometry directions (on the left-hand side) and the IIA ones with the $k$ isometry direction. These exotic fields correspond to some of the fields mentioned in \cite{BRKR}.
\begin{eqnarray}
\incl_l \mathcal{N}_7 &=& B_6 + 6 \incl_z B_6 \frac{ g_z}{ g_{zz}}\label{IIB_l} \\
\incl_k \incl_l N_7 &=& \incl_k \incl_z N_7 - 5 \incl_k \incl_z C_5 (\incl_k C_3 - 2 \incl_k \incl_z C_3 \frac{ g_z}{ g_{zz}} ) - 5 (\incl_k C_5 - 4 \incl_k \incl_z C_5 \frac{g_z}{ g_{zz}})\incl_k \incl_z C_3 \nonumber\\
\incl_k \incl_l  \mathcal{N}_8 &=& \incl_k  N_7 - 6 \incl_k \incl_z  N_7 \frac{ g_z}{ g_{zz}}\nonumber \\
\incl_k \incl_l  N_8 &=& \incl_k \incl_z  N_8 - 6 \incl_k \incl_z  C_3 (\incl_k  B_6 + 5 \incl_k \incl_z  B_6 \frac{g_z}{ g_{zz}}) - 15 (\incl_k C_3 + 2 \incl_k \incl_z C_3 \frac{ g_z}{ g_{zz}})\incl_k \incl_z B_6 \nonumber \\ & & + 20 \incl_k \incl_z  C_3( C_3 - 3 \incl_z  C_3 \frac{ g_z}{ g_{zz}})(\incl_k  C_3 - 2 \incl_k \incl_z  C_3 \frac{ g_z}{ g_{zz}}) + 30 \incl_z  C_3 (\incl_k  C_3 - 2 \incl_k \incl_z  C_3 \frac{ g_z}{ g_{zz}})^2 \nonumber \\
\incl_k \incl_l  \mathcal{N}_9 &=& \incl_k  N_8 + 7 \incl_k \incl_z  N_8 \frac{ g_z}{ g_{zz}}\nonumber \\
\incl_k \incl_l  N_9 &=& \incl_k \incl_z  N_9 - 7 \incl_k \incl_z  C_3(\incl_k  N_7 - 6\incl_k \incl_z  N_7 \frac{g_z}{g_{zz}}) - 35 \incl_k \incl_z  C_5 (\incl_k C_3 - 2 \incl_k \incl_z C_3 \frac{g_z}{g_{zz}})^2\nonumber \\
\incl_k \incl_l  \mathcal{N}_{10} &=& \incl_k  N_9 - 8 \incl_k \incl_z  N_9 \frac{ g_z}{ g_{zz}}\nonumber \\
\incl_k \incl_l  N_{10} &=& \incl_k \incl_z N_{10} - 8 \incl_k \incl_z  C_3 (\incl_k  N_8 + 7 \incl_k \incl_z  N_8 \frac{g_z}{g_{zz}})- 28 (\incl_k C_3 - 2 \incl_k \incl_z  C_3 \frac{g_z}{g_{zz}}) \incl_k \incl_z N_8 \nonumber \\ & & + 168 \incl_k \incl_z C_3(\incl_k C_3 - 2\incl_k \incl_z C_3 \frac{g_z}{g_{zz}})(\incl_k B_6 - 5 \incl_k \incl_z B_6 \frac{g_z}{g_{zz}})\nonumber \\ & & + 210 (\incl_k  C_3 - 2 \incl_k \incl_z  C_3 \frac{g_z}{g_{zz}})^2\incl_k \incl_z B_6 - 210 \incl_k \incl_z  C_3 (\incl_k  C_3 - 2 \incl_k \incl_z C_3 \frac{g_z}{g_{zz}})^2( C_3 - 3 \incl_z  C_3\frac{g_z}{g_{zz}}) \nonumber \\ & & - 315 (\incl_k  C_3 - 2 \incl_k \incl_z C_3\frac{g_z}{g_{zz}})^3 \incl_z  C_3.\nonumber
\end{eqnarray}

As the last T-duality list these are the rules for the IIB exotic potentials dualized along the $k$ direction. These are quite similar to the previous ones, as expected, but some of the fields are interchanged or have another sign.
\begin{eqnarray}
\incl_l \incl_k \mathcal{N}_7 &=& \incl_l \incl_z N_7 - 5 \incl_l \incl_z  C_5 (\incl_l  C_3 - 2 \incl_l \incl_z  C_3 \frac{ g_z}{ g_{zz}} ) - 5 (\incl_l  C_5 - 4 \incl_l \incl_z  C_5 \frac{ g_z}{ g_{zz}})\incl_l \incl_z  C_3 \label{IIB_k} \\
\incl_k N_7 &=&  B_6 + 6 \incl_z B_6 \frac{ g_z}{ g_{zz}}\nonumber \\
- \incl_l \incl_k  \mathcal{N}_8 &=& \incl_l  N_7 - 6 \incl_l \incl_z  N_7 \frac{ g_z}{ g_{zz}}\nonumber \\
- \incl_l \incl_k  N_8 &=&  \incl_l \incl_z  N_8 - 6 \incl_l \incl_z  C_3 (\incl_l  B_6 + 5 \incl_l \incl_z  B_6 \frac{ g_z}{ g_{zz}}) -15 (\incl_l C_3 - 2 \incl_l\incl_z C_3 \frac{ g_z}{ g_{zz}})\incl_l\incl_z B_6 \nonumber \\ & & + 20 \incl_l \incl_z  C_3( C_3 - 3\incl_z  C_3 \frac{ g_z}{ g_{zz}})(\incl_l  C_3 - 2 \incl_l \incl_z  C_3 \frac{ g_z}{ g_{zz}}) + 30 \incl_z  C_3 (\incl_l  C_3 - 2 \incl_l \incl_z  C_3 \frac{ g_z}{ g_{zz}})^2 \nonumber \\
- \incl_l \incl_k  \mathcal{N}_9 &=& \incl_l \incl_z  N_9 - 7 \incl_l \incl_z  C_3(\incl_l  N_7 - 6\incl_l \incl_z  N_7 \frac{ g_z}{ g_{zz}}) - 35 \incl_l \incl_z  C_5 (\incl_l  C_3 - 2 \incl_l \incl_z  C_3 \frac{ g_z}{ g_{zz}})^2\nonumber \\
- \incl_l \incl_k  N_9 &=& \incl_l  N_8 + 7 \incl_l \incl_z  N_8 \frac{ g_z}{ g_{zz}}\nonumber \\
\incl_l \incl_k  \mathcal{N}_{10} &=& \incl_l  N_9 - 8 \incl_l \incl_z  N_9 \frac{ g_z}{ g_{zz}}\nonumber \\
\incl_l \incl_k  N_{10} &=& \incl_l \incl_z  N_{10} - 8 \incl_l \incl_z  C_3 (\incl_l  N_8 + 7 \incl_l \incl_z  N_8 \frac{ g_z}{ g_{zz}})- 28 (\incl_l  C_3 - 2 \incl_l \incl_z  C_3 \frac{ g_z}{ g_{zz}}) \incl_l \incl_z  N_8 \nonumber \\ & & + 168 \incl_l \incl_z  C_3(\incl_l C_3 - 2 \incl_l \incl_z  C_3 \frac{ g_z}{ g_{zz}})(\incl_l  B_6 - 5 \incl_l\incl_z B_6\frac{ g_z}{ g_{zz}}) \nonumber \\ & & + 210 (\incl_l  C_3 - 2 \incl_l \incl_z  C_3 \frac{ g_z}{ g_{zz}})^2 \incl_l \incl_z  B_6 - 210 \incl_l \incl_z  C_3 ( C_3 - 3 \incl_z  C_3\frac{ g_z}{ g_{zz}})(\incl_l  C_3 - 2\incl_l\incl_z C_3 \frac{ g_z}{ g_{zz}})^2 \nonumber \\ & & - 315 (\incl_l  C_3 - 2 \incl_l \incl_z C_3\frac{ g_z}{ g_{zz}})^3 \incl_z  C_3.\nonumber
\end{eqnarray}

\sect{Gauge variations}\label{sectgaugevar}

Gauge variations and coordinate transformation in the eleven-dimensional theory:
\begin{eqnarray}
\delta \hat k^{-2} \hat k_1 &=& \partial \hat \Lambda_0  \label{gv_11D}\\
\delta \hat C_3 &=& 3\partial \hat \Lambda_2 \nonumber \\
\delta \hat C_6 &=& 6\partial \hat \Lambda_5 + 30 \hat C_3 \partial \hat \Lambda_2 \nonumber \\ 
\delta \incl_k \hat C_8 &=& - 7 \partial (\incl_k \hat \Lambda_7) - 105 \incl_k \hat C_3 \partial (\incl_k \hat  \Lambda_5)  + 210 (\incl_k \hat C_3)^2 \partial \hat \Lambda_2 + 140 \incl_k \hat C_3 \hat C_3 \partial (\incl_k \hat \Lambda_2) \nonumber \\
\delta (\incl_k \hat C_{10}) &=& - 9\partial (\incl_k \hat \Lambda_9) - 252 \incl_k \hat C_3 \partial (\incl_k \hat \Lambda_7) - 1890 (\incl_k \hat C_3)^2 \partial (\incl_k \hat \Lambda_5) \nonumber \\
& & + 2835 (\incl_k \hat C_3)^3 \partial \hat \Lambda_2 + 1890 \hat C_3 (\incl_k \hat C_3)^2 \partial (i_k \hat \Lambda_2)\nonumber.
\end{eqnarray}

Gauge variations in Type IIA:
\begin{eqnarray}
\delta  k^{-2} k_1 &=& \partial  \xi_0 \label{gvIIA} \\
\delta  B_2 &=& 2\partial  \Sigma_1 \nonumber \\
\delta  C_{2n+1} &=& \sum_{p=0}^{n} \frac{(2n +1)!}{2^p p! (2n-2p)!} ( B_2)^p \partial\Lambda_{2n-2p} \nonumber \\
\delta  B_6 &=& 6 \partial  \Sigma_5 + 30  C_3 \partial  \Lambda_2 - 6  C_5\partial \Lambda_0 + 30  C_3  B_2  \partial  \Lambda_0 \nonumber \\
\delta \incl_k  N_7 &=& - 6 \partial \incl_k  \xi_6 - 30 \incl_k  B_2 \partial (\incl_k  \Sigma_5) - 60 \incl_k  C_3 \partial (\incl_k  \Lambda_4) - 180 \incl_k  C_3  B_2 \partial (\incl_k  \Lambda_2) + 180 \incl_k  C_3 \incl_k  B_2 \partial  \Lambda_2 \nonumber \\ & &  + 180 \incl_k  C_3  B_2 \incl_k  B_2 \partial  \Lambda_0 \nonumber \\
\delta \incl_k  N_8 &=& - 7 \partial (\incl_k \xi_7) - 105 \incl_k  C_3 \partial (\incl_k  \Sigma_5) + 140 \incl_k  C_3  C_3 \partial (\incl_k  \Lambda_2) + 210 \incl_k  C_3 \incl_k  C_3 \partial \Lambda_2 \nonumber \\ & & - 7 \incl_k  N_7 \partial  \Lambda_0 - 140 \incl_k  C_3  C_3 \incl_k  B_2 \partial  \Lambda_0 + 210 \incl_k  C_3 \incl_k  C_3  B_2 \partial  \Lambda_0 \nonumber \\
\delta \incl_k  N_9 &=& - 8 \partial (\incl_k  \xi_8)- 56 \incl_k  B_2 \partial (\incl_k  \xi_7) - 168 \incl_k  C_3 \partial (\incl_k  \xi_6) - 840 \incl_k  C_3 \incl_k C_3 \partial (\incl_k  \Lambda_4) \nonumber \\ & & - 840 \incl_k  C_3 \incl_k B_2 \partial (\incl_k  \Sigma_5) + 2520 (\incl_k C_3)^2 \incl_k B_2 \partial \Lambda_2 \nonumber \\ & & - 2520  B_2 (\incl_k C_3)^2 \partial (\incl_k  \Lambda_2) + 2520 (\incl_k C_3)^2 B_2 \incl_k B_2 \partial \Lambda_0 \nonumber \\
\delta \incl_k  N_{10} &=& -9 \partial (\incl_k  \zeta_9) - 252 \incl_k  C_3 \partial (i_k  \xi_7) - 1890 (\incl_k  C_3)^2 \partial (\incl_k  \Sigma_5) + 2835  (\incl_k C_3)^3 \partial \Lambda_2 \nonumber \\ & & + 1890  C_3 (\incl_k C_3)^2 \partial (\incl_k  \Lambda_2) - 9 \incl_k  N_9\partial \Lambda_0 \nonumber \\ & & 
+ 2835 (\incl_k  C_3)^3 B_2\partial\Lambda_0 - 1890 C_3 (\incl_k C_3)^2 i_k B_2 \partial \Lambda_0.\nonumber
\end{eqnarray}

Gauge variations in Type IIB:
\begin{eqnarray}
\delta B_2 &=& 2\partial \Sigma_1 \label{gvIIB} \\
\delta C_{2n} &=& \sum_{p=0}^{n-1} \frac{(2n)!}{2^p p!(2n-2p-1)!} B_2^p \partial\Lambda_{2n-2p-1} \nonumber \\
\delta B_6 &=& 6\partial \Sigma_5 - 30 C_4 \partial \Lambda_1 \nonumber \\
\delta Q_8 &=& 8 \partial \lambda_7 - 56 B_6 \partial \Lambda_1 \nonumber \\
\delta Q_{10} &=& 10 \partial \lambda_9 - 90 Q_8 \partial \Lambda_1 \nonumber \\
& & \nonumber \\
\delta \incl_l \mathcal{N}_7 &=& - 6 \partial \incl_l \zeta_6 - 30 \incl_l C_4 \partial (\incl_l \Lambda_3) + 6 \incl_l  C_6 \partial (\incl_l  \Lambda_1)-30 \incl_l C_4 B_2 \partial (\incl_l \Lambda_1) - 30 \incl_l B_2 \partial (\incl_l  \Sigma_5)\nonumber \\ & & + 60 \incl_l C_4 \incl_l B_2 \partial \Lambda_1 \nonumber \\
\delta \incl_k N_7 &=& - 6 \partial \incl_k \xi_6 - 30 \incl_k C_4 \partial (\incl_k  \Lambda_3) + 6 \incl_k  C_6 \partial (\incl_k  \Lambda_1) - 30 \incl_k  C_4 B_2 \partial (\incl_k \Lambda_1) - 30 \incl_k  B_2 \partial (\incl_k  \Sigma_5)\nonumber \\ & & + 60 \incl_k  C_4 \incl_k  B_2 \partial \Lambda_1 \nonumber \\
\delta (\incl_k \incl_l \mathcal{N}_8) &=& 6 \partial (\incl_k \incl_l  \zeta_7) + 30 \incl_k  B_2 \partial (\incl_k \incl_l \zeta_6) - 30 \incl_l B_2 \partial (\incl_k \incl_l \xi_6) - 120 \incl_k B_2 \incl_l B_2 \partial (\incl_k \incl_l  \Sigma_5) \nonumber \\ & & + 60 \incl_k \incl_l  C_4 \partial (\incl_k \incl_l  \Lambda_5) + 180 \incl_k \incl_l C_4 \incl_l B_2 \partial \incl_k \Lambda_3 - 180 \incl_k \incl_l  C_4 \incl_k B_2 \partial (\incl_l  \Lambda_3)  \nonumber \\ & & + 180 \incl_k \incl_l  C_4 B_2 \partial (\incl_k \incl_l \Lambda_3) - 180 \incl_k \incl_l  C_4 B_2 \incl_k  B_2 \partial (i_l  \Lambda_1) +180 \incl_k \incl_l C_4 B_2 \incl_l  B_2 \partial  (\incl_k \Lambda_1)  \nonumber \\ & &  +360 \incl_k \incl_l C_4 \incl_k B_2 \incl_l  B_2 \partial \Lambda_1\nonumber \\ 
\delta (\incl_k \incl_l N_8) &=& 6 \partial (\incl_k \incl_l \xi_7) - 30 \incl_k \incl_l B_6 \partial (\incl_k \incl_l \Lambda_3) - 6\incl_k \incl_l \mathcal{N}_7 \partial (\incl_k \Lambda_1) + 6 \incl_k \incl_l N_7 \partial (\incl_l \Lambda_1) \nonumber \\ & & -30 \incl_k \incl_l B_6 \incl_l B_2 \partial (\incl_k \Lambda_1) + 30 \incl_k \incl_l B_6 \incl_k B_2 \partial (\incl_l \Lambda_1) \nonumber \\ & & 
+ 30 \incl_k \incl_l C_4 \incl_l C_4 \partial (\incl_k \Lambda_1) -30 \incl_k \incl_l C_4 \incl_k C_4 \partial (\incl_l \Lambda_1) + 90 (\incl_k \incl_l C_4)^2 \partial \Lambda_1 \nonumber \\
\delta (\incl_k \incl_l  \mathcal{N}_9) &=& 7 \partial (\incl_k \incl_l \zeta_8) - 42 \incl_l  B_2 \partial (\incl_k \incl_l \xi_7) + 105 \incl_k \incl_l C_4 \partial(\incl_k \incl_l \zeta_6) - 420 \incl_k \incl_l C_4 \incl_l B_2 \partial(\incl_k \incl_l \Sigma_5) \nonumber \\ & & - 140 \incl_k \incl_l C_4 \incl_l C_4 \partial (\incl_k \incl_l  \Lambda_3) - 210 (\incl_k \incl_l C_4)^2 \partial (\incl_l  \Lambda_3) \nonumber \\ & & + 7 \incl_k \incl_l \mathcal{N}_8\partial (\incl_l\Lambda_1) + 140 \incl_k \incl_l C_4 \incl_l C_4 \incl_k B_2 \partial(\incl_l  \Lambda_1)\nonumber \\ & & - 140 \incl_k \incl_l C_4 \incl_l C_4 \incl_l B_2 \partial(\incl_k  \Lambda_1) -210 (\incl_k \incl_l C_4)^2 B_2 \partial (\incl_l  \Lambda_1)+ 420 (\incl_k \incl_l C_4)^2 \incl_l B_2 \partial \Lambda_1 \nonumber \\
\delta (\incl_k \incl_l  N_9) &=& 7 \partial (\incl_k \incl_l \xi_8) - 42 \incl_k  B_2 \partial (\incl_k \incl_l \xi_7) + 105 \incl_k \incl_l C_4 \partial(\incl_k \incl_l \xi_6) - 420 \incl_k \incl_l C_4 \incl_k B_2 \partial(\incl_k \incl_l \Sigma_5) \nonumber \\ & & - 140 \incl_k \incl_l C_4 \incl_k C_4 \partial (\incl_k \incl_l \Lambda_3) - 210 (\incl_k \incl_l C_4)^2 \partial (\incl_k \Lambda_3) \nonumber \\ & & + 7 \incl_k \incl_l \mathcal{N}_8\partial (\incl_k\Lambda_1) + 140 \incl_k \incl_l C_4 \incl_k C_4 \incl_k B_2 \partial(\incl_l \Lambda_1)\nonumber \\ & & - 140 \incl_k \incl_l C_4 \incl_k C_4 \incl_l B_2 \partial(\incl_k  \Lambda_1) - 210 (\incl_k \incl_l C_4)^2 B_2 \partial (\incl_k  \Lambda_1) + 420 (\incl_k \incl_l C_4)^2 \incl_k B_2 \partial \Lambda_1 \nonumber \\
\delta (\incl_k \incl_l  \mathcal{N}_{10}) &=& 8\partial(\incl_k \incl_l \xi_9) - 56 \incl_l  B_2 \partial (\incl_k \incl_l \zeta_8) + 56 \incl_k  B_2 \partial (\incl_k \incl_l \xi_8) - 336 \incl_k B_2 \incl_l B_2 \partial (\incl_k \incl_l  \xi_7) \nonumber \\ & & + 168 \incl_k \incl_l  C_4 \partial (\incl_k \incl_l \zeta_7) + 840 \incl_k \incl_l C_4 \incl_k B_2 \partial (\incl_k \incl_l \zeta_6) - 840 \incl_k \incl_l C_4 \incl_l B_2 \partial (\incl_k \incl_l \xi_6) \nonumber \\ & & + 840 (\incl_k \incl_l C_4)^2 \partial (\incl_k \incl_l \Lambda_5) -3360 \incl_k \incl_l C_4 \incl_k B_2\incl_l B_2 \partial (\incl_k \incl_l \Sigma_5) + 2520 (\incl_k \incl_l C_4)^2 \incl_l  B_2 \partial (\incl_k \Lambda_3) \nonumber \\ & & - 2520 (\incl_k \incl_l C_4)^2 \incl_k B_2 \partial (\incl_l \Lambda_3) + 2520 (\incl_k \incl_l  C_4)^2 B_2 \partial (\incl_k \incl_l \Lambda_3) \nonumber \\ & & + 5040 (\incl_k \incl_l C_4)^2 \incl_k B_2 \incl_l B_2 \partial \Lambda_1 + 2520 (\incl_k \incl_l C_4)^2 B_2 \incl_l B_2 \partial (\incl_k \Lambda_1)\nonumber \\ & & - 2520 (\incl_k \incl_l C_4)^2 B_2 \incl_k B_2 \partial (\incl_l  \Lambda_1)\nonumber \\
\delta  (\incl_k \incl_l N_{10}) &=& 8 \partial (\incl_k \incl_l  \zeta_9) - 56 \incl_k \incl_l N_8 \partial (\incl_k \incl_l \Lambda_3) - 56 \incl_k \incl_l N_8 \incl_l B_2 \partial (\incl_k  \Lambda_1) + 56 \incl_k \incl_l N_8 \incl_k B_2 \partial (\incl_l \Lambda_1) \nonumber \\ & & - 8 \incl_k \incl_l \mathcal{N}_9 \partial (\incl_k \Lambda_1) + 8 \incl_k \incl_l N_9 \partial (\incl_l  \Lambda_1) - 168 \incl_k \incl_l N_7 \incl_k \incl_l C_4 \partial (\incl_l \Lambda_1)\nonumber \\ & & + 168 \incl_k \incl_l \mathcal{N}_7 \incl_k \incl_l C_4 \partial (\incl_k \Lambda_1) - 280 \incl_l C_4 (\incl_k \incl_l  C_4)^2 \partial (\incl_k  \Lambda_1) \nonumber \\ & & - 280 \incl_k  C_4 (\incl_k \incl_l C_4)^2 \partial (\incl_l \Lambda_1) - 840 (\incl_k \incl_l C_4)^3 \partial \Lambda_1.\nonumber
\end{eqnarray}

\end{document}